\documentclass[%
reprint,
superscriptaddress,
showpacs,
amsmath,amssymb,amsfonts,
aps,
prx,
floatfix,notitlepage,latexsym
]{revtex4-1}

\usepackage[dvipdfmx]{graphicx}
\usepackage{dcolumn}
\usepackage{bm}
\usepackage{braket}
\usepackage{mediabb} 
\usepackage[dvipdfmx]{attachfile2}

\begin{abstract}
Correlated many-body problems ubiquitously appear in various fields of physics such as condensed matter physics, nuclear physics, and statistical physics. However, due to the interplay of the large number of degrees of freedom, it is generically impossible to treat these problems from first principles. Thus the construction of a proper model, namely effective Hamiltonian, is essential.
Here, we propose a simple supervised learning algorithm for constructing Hamiltonians from given energy or entanglement spectra. We apply the proposed scheme to the Hubbard model at the half-filling, and compare the obtained effective low-energy spin model with several analytic results based on the high order perturbation theory which have been inconsistent with each other. We also show that our approach can be used to construct the entanglement Hamiltonian of a quantum many-body state from its entanglement spectrum as well.  We exemplify this using the ground states of the $S=1/2$ two-leg Heisenberg ladders. We observe a qualitative difference between the entanglement Hamiltonians of the two phases (the Haldane phase and the Rung Singlet phase) of the model due to the different origin of the entanglement. In the Haldane phase, we find that the entanglement Hamiltonian is non-local by nature, and the locality can be restored by introducing the anisotropy and turning the ground state into the large-$D$ phase.  Possible applications to the model construction from experimental data and to various problems of strongly correlated systems are discussed.

\vspace{3mm}
\noindent PhySH: Optimization problems, Machine learning, Strongly correlated systems, Quantum entanglement
\end{abstract}
\begin{document}


\title{Construction of Hamiltonians by supervised learning of energy and entanglement spectra}
\author{Hiroyuki Fujita}
\thanks{Corresponding author}
\email{h-fujita@issp.u-tokyo.ac.jp}
\author{Yuya O. Nakagawa}
\author{Sho Sugiura}
\author{Masaki Oshikawa}
\affiliation{Institute for Solid State Physics, The University of Tokyo. Kashiwa, Chiba 277-8581, Japan}
\date{\today}

\maketitle

\section{Introduction}

Physical properties of classical or quantum systems are
completely determined by their Hamiltonians.
Therefore, the construction of the Hamiltonian for a system of interest
is at the heart of physics.
Traditionally, the construction has been mostly performed in an empirical manner; one would write down a simple model Hamiltonian with a few
tunable parameters, and then fit them to match experimental data. These days, our knowledge of the microscopic details of the system
often allows a quantitatively accurate construction of
the Hamiltonian of particular materials from first principles~\cite{:2008aa,:2009aa,PhysRevB.80.155134,PhysRevLett.113.107201}. However, the construction of the (effective) Hamiltonian appears also
in many other contexts, and a more general and systematic method for a given microscopic model and/or experimental data is highly desired.

The construction of the Hamiltonian may be regarded as
a particular case of \emph{modeling} in the data science. This viewpoint opens up many new possibilities~\cite{:2015aa,Honma:2016aa,otsuki2017,hukushima_2017,PhysRevB.95.064407} to understand and exploit various data of physical systems. In particular, one can use modern tools of the data science like the machine learning (ML). Specifically speaking, we could develop a ML algorithm which automatically determines the optimized model for the given data. Such
a ML approach has many advantages. For one thing, it would suffer less
from the bias of each researcher or the bias can be easily controlled,
because the so-called cost function gives a standard for quantifying the degree of the success of the modeling. For another, the ML can treat problems which are difficult for human
beings due to, for example, the large number of
parameters or the absence of mathematically tractable methods. Although applications of the ML to physics and materials science have been actively explored~\cite{Kusne:2014aa,PhysRevLett.114.105503,Kalinin:2015aa,Roger_ML_2,Melko_ML_1,doi:10.7566/JPSJ.85.123706,YiZhang2016,D.Sarma2016,Schndler2017,Carleo602,Nieuwenburg:2017aa,Carrasquilla:2017aa,YiZhang2017} recently, these existing studies are mainly aimed to find new functional materials or to solve many-body problems (detection of the phase
transitions for example), and the application for the construction of Hamiltonians is unexplored.

In this paper, we propose a scheme of reverse engineering the Hamiltonian from its low-lying (energy or entanglement) spectrum using the ML. Specifically speaking, we formulate the construction of Hamiltonian as a supervised learning problem for given energy or entanglement spectrum.
This is interesting in several respects.  First, this gives a systematic
method to construct a low-energy effective Hamiltonian with a reduced
number of degrees of freedom, allowing us to perform various numerical calculations of large systems. Second, it can deepen our understanding of quantum many-body systems through the explicit construction of their Entanglement (Modular)
Hamiltonian (EH) using its spectrum, namely the Entanglement Spectrum
(ES)~\cite{PhysRevLett.101.010504}.

 We show that combining the gradient descent method of the ML and the basic perturbation theory of quantum mechanics, we can update the parameters of a trial Hamiltonian securely, which eventually leads to a (local) minimum of the cost function. We first apply the scheme for the construction of the effective spin Hamiltonians of the Hubbard models at the half-filling. That is, we use the low-energy spectrum of the Hubbard model as the ``experimental'' data, and construct the effective spin Hamiltonian. Taking the homogeneous Hubbard  chain as an example, we quantitatively study the accuracy of the ML estimation. Although the effective spin model for the Hubbard chain has been constructed in the perturbation theory~\cite{PhysRevB.8.2236,Takahashi,PhysRevB.37.9753,Rej2006}, higher-order calculations are rather complicated. In fact, conflicting results have been reported in the literature.
 
 We resolve the controversy in the sixth order perturbation terms among previous works~\cite{PhysRevB.8.2236,PhysRevB.37.9753,Rej2006}. This demonstrates the accuracy and the practical utility of our approach. We also show that even when the original Hubbard model is not translationally invariant, i.e. spatially modulated, we can optimize the local Hamiltonian at each site independently to find the effective model. In other words, using ML we can readout the symmetry breaking only from the low-energy spectrum of the original model.

We also demonstrate the construction of the EH from the ES of a given quantum many-body state. As many studies suggest, there is a similarity between the ES and the physical energy spectrum of the subsystem, called the entanglement/edge correspondence~\cite{PhysRevLett.101.010504,0295-5075-96-5-50006,PhysRevB.84.205136,PhysRevLett.108.196402,PhysRevB.86.045117,PhysRevB.86.245310}. Using the two-leg Heisenberg ladder~\cite{PhysRevLett.105.077202,PhysRevB.85.054403,PhysRevB.88.245137,PhysRevB.83.245134} as an example, we show that the ML gives an estimate for the explicit lattice EH through the optimization of the spectrum. Away from the strongly antiferromagnetic rung limit, the EH is in general non-local as was argued in the earlier study~\cite{PhysRevB.83.245134} using the full-diagonalization of the reduced density matrix. Compared to the full-diagonalization approach, the ML method is based only on the low-lying spectrum, so that its computational cost is much cheaper and potentially we can access very large system sizes. In addition to the two-body Heisenberg interactions which have been well studied in the previous works, we examine the role of the four-body interactions in the EH using our ML scheme. Then, we observe a qualitative difference between the EH of the rung singlet phases and the Haldane phase, which goes beyond the  field-theoretical understandings~\cite{PhysRevB.88.245137}. We show that the ES in the Haldane phase is intrinsically non-local and the entanglement/edge correspondence is broken there, while we can restore that with the help of the SU(2) breaking anisotropy.
 
 The rest of this paper is organized as follows.  In Sec.~\ref{Sec. ML algorithm}, we give a brief review of the gradient descent method and propose its implementation for the construction of Hamiltonians. In Sec.~\ref{Sec. Benchmark}, as a benchmark of the algorithm presented in Sec.~\ref{Sec. ML algorithm}, we derive the effective spin model of the Hubbard models. In Sec.~\ref{HA construction}, as a more nontrivial application, we construct the EH of the two-leg Heisenberg ladder using the ES of its ground state. The final section is devoted to the conclusion and outlook, where we discuss possible applications of the presented method.
  
\section{Gradient descent method for energy spectrum}\label{Sec. ML algorithm}
Here, we quickly review the basic ML algorithm, the gradient descent method, which treats the following problem: given data $\{y_i\}$ ($i = 0,...,N-1$) of our interest and explanatory variables $\{ c_j\}$ ($j = 1,...,M$), we identify the relation between them. Specifically speaking, we set a hypothesis like $\tilde{y}_i = f_i(c_j)$ as a function of the parameters $\{c_j\}$ and define a cost function ${\rm Cost}(\{c_j\}) =\frac{1}{2N}\sum_{i=0}^{N-1} (y_i - \tilde{y_i})^2$. This cost function measures how well the hypothesis explains the data $\{y_i\}$ for the given parameters $\{c_j\}$. In the gradient descent method, we update the parameters by using the gradient of the cost function as 
$c_j \rightarrow c_j - \alpha \frac{\partial {\rm Cost}(\{ c_j\})}{\partial c_j}$, 
where $\alpha$ is a hyperparameter called the learning rate, which is chosen by hand. If $\alpha$ is too large, there is no guarantee that the cost decreases at each step, while if it is too small the optimization proceeds quite slowly. As long as the learning rate $\alpha$ is properly chosen (at least not too large), the parameter update eventually leads to a (local) minimum of the cost function. 

In this paper, we construct a (spin) Hamiltonian which reproduces the given spectrum $y_i = E^{\rm data}_i $ ($i = 0,...,N-1$) as the $N$ lowest energy eigenvalues $\{ E_i\}$. The hypothesis $\tilde{y}_i$ is a function of coupling constants $\{c_j \}$ ($j = 1,...,M$) in the trial Hamiltonian $H = \sum_{j=1}^{M} c_j H_j$, where $H_j$ can be exchange interactions, coupling with external fields, and other multi-spin interactions. 

As we noted above, to apply the gradient descent method, it is essential to know the gradient of the cost function in terms of the coupling constants. However, the relation between the energy spectrum and the coupling constants in the Hamiltonian is quite complicated in general. Especially in correlated systems, the full parameter dependence of the cost function is in most cases unknown. Nevertheless, if the learning rate $\alpha$ is small, the gradient itself can be analytically calculated, by using the first order perturbation theory of quantum mechanics: for given energy eigenstates of the (spin) Hamiltonian $\ket{\Psi_i (\{c_j \})}$ and their eigenenergies $E_i (\{c_j \})$, we can update the parameter $c_j$ as
\begin{align} 
c_j \rightarrow \hspace{2mm}&c_j - \alpha \frac{\partial {\rm Cost}(\{ c_j\})}{\partial c_j} \\
&= c_j - \alpha\sum_{i=0}^{N-1} \frac{\partial {\rm Cost}(\{c_j\})}{\partial E_i(\{c_j \})} \frac{\partial E_i (\{c_j \})}{\partial c_j}\nonumber \\
&\simeq c_j - \alpha \sum_{i=0}^{N-1}  \frac{\partial {\rm Cost}(\{c_j\})}{\partial E_i(\{c_j \})} \braket{\Psi_i| H_j|\Psi_i }\nonumber.
\end{align}
In particular, if we use the mean squared error divided by two as the cost function
\begin{align}
{\rm Cost }(\{ c_j\}) =  \frac{1}{2N}\sum_{i = 0}^{N-1} \left[E^{\rm data}_i  - E_i (\{c_j \})\right]^2 \label{Cost_function},
\end{align}
we have the following update scheme for the parameters:
\begin{align}
c_j  \rightarrow c_j + \alpha \frac{1}{N}\sum_{i=0}^{N-1}  \left[E^{\rm data}_i -E_i (\{c_j \})\right] \braket{\Psi_i| H_j|\Psi_i }.\label{algorithm}
\end{align}
As long as the learning rate $\alpha$ is small, Eq.~\eqref{algorithm} ensures that the cost improves at each step, so we can find an optimized spin Hamiltonian whose spectrum achieves a (local) minimum of the cost function. Since the strategy here is so simple, it is compatible with various advanced algorithms developed in the context of supervised learning, such as sparse modeling~\cite{Honma:2016aa,hukushima_2017,otsuki2017} or stochastic optimization~\cite{Adam} as we show later (see also appendix A and B).

Since there is an infinite number of different Hamiltonians with exactly
the same eigenvalues, the posed problem, the construction of the Hamiltonian from a given spectrum, is undetermined by
nature (mathematically this corresponds to the degrees of freedom of the unitary transformations, i.e. the ambiguity of the basis).
However, what we are dealing with is not a featureless Hermitian matrix, but an (entanglement) Hamiltonian of some physical systems.
Therefore, we can physically set an ansatz for the spin Hamiltonian, which, for example, possesses some symmetries or the locality of interactions. Such assumptions allow us to reduce the number of the parameters to be optimized and narrow down the candidate of the physically plausible Hamiltonian within that ansatz.

\section{Low-energy Hamiltonian of Hubbard model} \label{Sec. Benchmark}
In the previous section, we presented a ML scheme for the construction of Hamiltonians from a given spectrum. In this section, we derive the effective spin models of the Hubbard models and compare them with the results from the perturbation theory~\cite{PhysRevB.8.2236,Takahashi,PhysRevB.37.9753,Rej2006}.

We consider the following fermionic Hubbard chain of length $L$ (periodic) with the spatially modulated repulsive interaction:
\begin{align}
H = \sum_{\substack{i = 1, ..., L \\\sigma = \uparrow,\downarrow}} t(c_{i+1,\sigma}^\dag c_{i,\sigma} + c_{i,\sigma}^\dag c_{i+1,\sigma}) + U_i n_i^\uparrow n_i^\downarrow \label{Hubbard}
\end{align}
where $U_i >0$ is the site-dependent on-site repulsive interaction, and $n_i^\sigma = c^\dag{}_{i, \sigma}c_{i, \sigma}$ ($\sigma = \uparrow, \downarrow$) is the electron density at site $i$ with spin $\sigma$.  Hereafter we focus on the half-filled case, so that the total number of electrons is $L$. We use the mean squared error Eq.~\eqref{Cost_function} as the cost function and take $t = 1$ in the following. For the optimization, we choose a particular total magnetization sector ($S^z= \sum_i (n^{\uparrow}_i - n^{\downarrow}_i)/2=\sum_i S^z_i  =2$ for Sec.~\ref{3-a} and $S^z = 4.5$ for Sec.~\ref{3-b}). As long as  the low-energy part in that sector is described as a spin model, the choice of the magnetization sector is arbitrary. One can choose the computationally cheap sector so long as the number of available eigenvalues are sufficient to perform the learning and cross-validation.

\subsection{Homogeneous Hubbard model: comparison with the perturbation theory}\label{3-a}
First, we consider the homogeneous Hubbard chain $U_i = U$ for all $i = 1, ..., L = 10$ where we can use the perturbation theory as the reference. As is well known from the perturbation theory, when the onsite repulsive interaction is dominantly strong $U/t \gg 1$ the low-energy physics of the model Eq.~\eqref{Hubbard} is captured with the nearest-neighbor antiferromagnetic Heisenberg model. 

As the ratio $U/t$ becomes smaller, other SU(2) symmetric terms such as further neighbor and/or multi-spin couplings start to manifest. From the ML perspective, the simplest strategy is to make the ansatz complex enough. Indeed, it is generically true that increasing the number of terms (complexity) in the ansatz can reduce the optimized cost evaluated for the dataset used for the optimization. However, increasing the complexity of the ansatz can cause the overfitting and the estimated Hamiltonian can give wrong predictions for physical quantities. Therefore, in order to quantify the success of the modeling, using the cost evaluated for the data not used in the optimization, namely the cross-validation, is important [below we use the lowest 40 eigenvalues in the $S^z = 3$ sector as the cross-validation set and calculate the mean-squared error for them]. 

\begin{figure}[htbp]
   \centering
\includegraphics[width = 85mm]{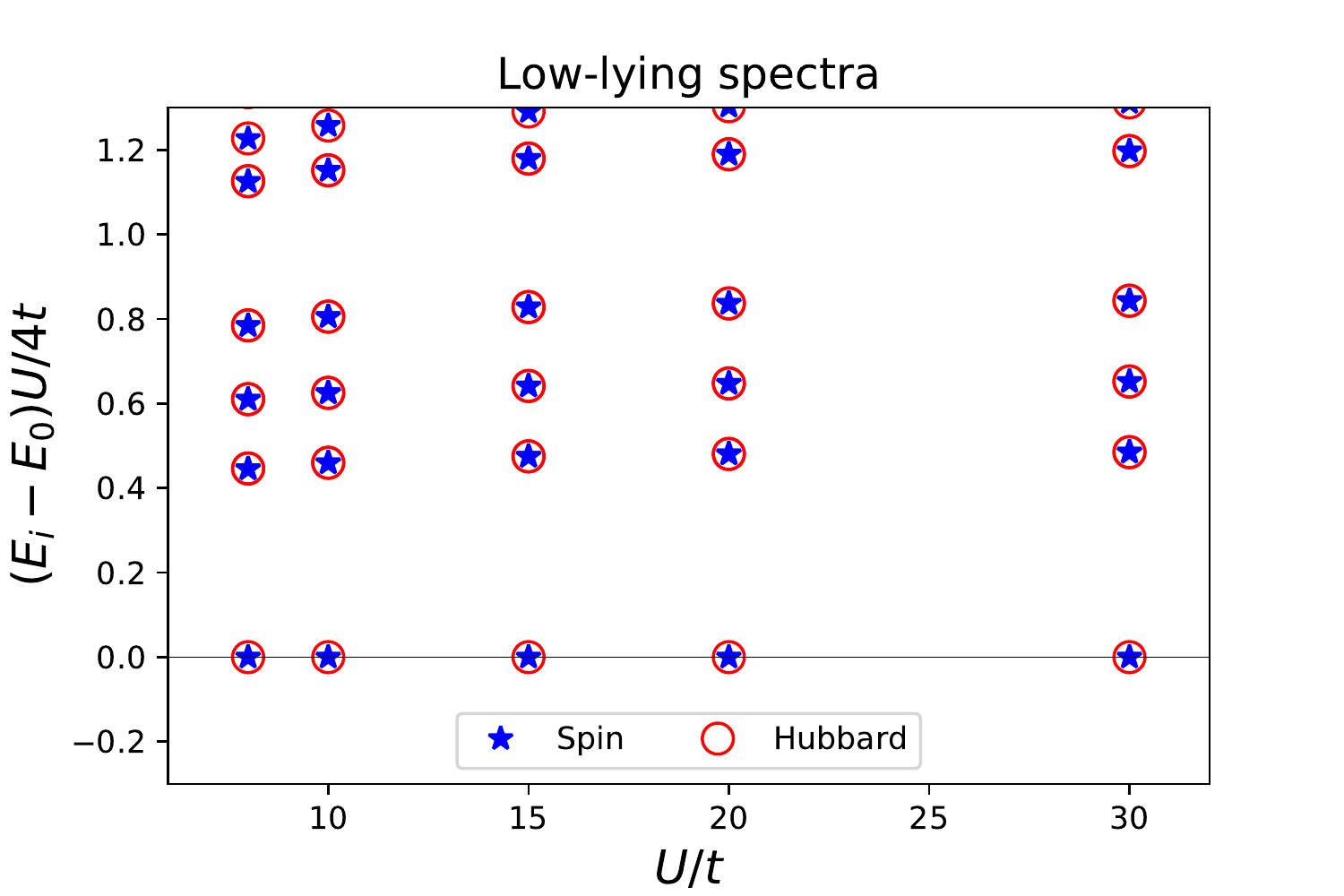}
       \caption{Direct comparison of the spectrum (in the total magnetization $S^z = 2$ sector) of the Hubbard model (red circles) with length $L=10$ and of the estimated spin model (blue stars) under the ansatz Eq.~\eqref{O5_ANSATZ}. The estimated spin Hamiltonian well reproduces the low-lying spectrum of the Hubbard model. We use the lowest $N=50$ eigenvalues in the $S^z = 2$ sector for optimization with the learning rate $\alpha = 0.1$. }
       \label{Spectra_residue}
  \end{figure}

As the ansatz for the effective model, here we consider the following $O(t^5/U^5)$ Hamiltonian determined through the sparse modeling approach (the $L_1$ norm regularization; see appendix A):

\begin{align}
H_{\rm spin} &= E_c+ \sum_{i=1}^L\sum_{\Delta = 1,2,3} J_\Delta \bm{S}_{i+\Delta} \cdot \bm{S}_i  \nonumber \\
&+ \sum_{i=1}^L K_2 (\bm{S}_{i} \cdot \bm{S}_{i+2} )(\bm{S}_{i+1} \cdot \bm{S}_{i+3} )\nonumber \\ &+  \sum_{i=1}^L K_3 (\bm{S}_{i} \cdot \bm{S}_{i+3} )(\bm{S}_{i+1} \cdot \bm{S}_{i+2} ),
\label{O5_ANSATZ}
\end{align}
where $E_c$, $J_\Delta$, and $K_\alpha$ are parameters to be determined. The constant term $E_c$ can be fixed by equating the average of the input spectrum to that of the ansatz.
In Fig.~\ref{Spectra_residue}, we directly compare the low-lying spectra of the Hubbard model and the estimated spin model for several values of $U$.  It shows that the estimated Hamiltonian under the ansatz Eq.~\eqref{O5_ANSATZ} well reproduces the spectrum of the Hubbard model. We also notice that at large $U$, where the higher order terms of the perturbation theory are inessential, the normalized spectrum becomes independent of $U/t$, since the spectrum is converged to that of the nearest neighbor Heisenberg model with  $J_1 = 1$.

Next, we compare the obtained values of the couplings with those from the perturbation theory. Although the Hubbard chain is such a simple system, performing the high order perturbative calculations correctly is hard. As a result, there are several inconsistent reports for the $O(t^5/U^5)$ terms~\cite{PhysRevB.8.2236,PhysRevB.37.9753,Rej2006}. Here we examine the estimation accuracy of the ML approach for the ansatz Eq.~\eqref{O5_ANSATZ} quantitatively and show that the one presented in Ref.~\onlinecite{PhysRevB.37.9753} is the correct one:

\begin{align}
\frac{J^{ p}_1}{t} &= \frac{4t}{U} - \frac{16 t^3}{U^3} + \frac{116 t^5}{U^5}  + O\left(\frac{t^7}{U^7}\right),\\
\frac{J^{ p}_2}{t}&=\frac{4t^3}{U^3} - \frac{40 t^5}{U^5}  + O\left(\frac{t^7}{U^7}\right),\\
\frac{J^{ p}_3}{t} &= \frac{4t^5}{U^5}  +O\left(\frac{t^7}{U^7}\right), \\
\frac{K^{ p}_2}{t}&=-\frac{16t^5}{U^5}  + O\left(\frac{t^7}{U^7}\right),\\
\frac{K^{ p}_3}{t} &= \frac{16t^5}{U^5}  +O\left(\frac{t^7}{U^7}\right).
\label{MCD}
\end{align}
Here we use the subscript $p$ to indicate that these are obtained from the perturbation theory.

\begin{figure}[htbp]
   \centering
\includegraphics[width = 90mm]{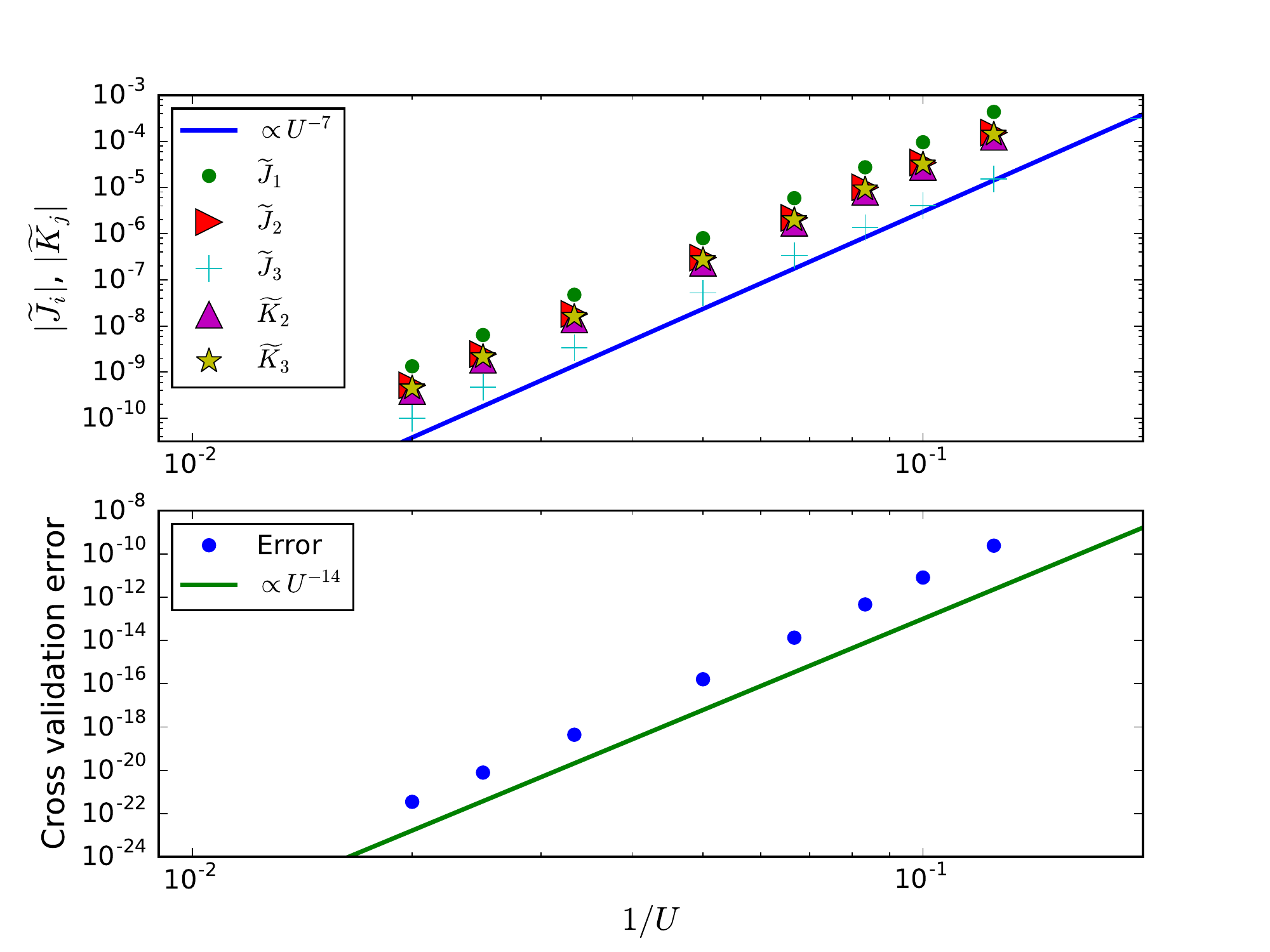}
       \caption{Upper panel: Comparison of couplings obtained by the perturbation theory $J^{\rm p}_{i=1,2,3}$, $K^{\rm p}_{j=2,3}$ in Ref.~\onlinecite{PhysRevB.37.9753} and by the optimization scheme Eq.~\eqref{algorithm}. For all the couplings their differences $\widetilde{J}_i = J_i - J^p_i, \widetilde{K}_j = K_j - K^p_j$ depend on $U$ as $U^{-7}$ for large $U$, indicating that the proposed scheme Eq.~\eqref{algorithm} can reproduce the perturbative results. Lower panel: $U$ dependence of the cross-validation error: $\frac{1}{40}\sum_{i=0}^{39}(E^{\rm Hubbard}_i - E^{\rm spin}_i)^2$ evaluated for eigenvalues in the $S^z = 3$ sector. The cost, mean squared error of the estimation, depends on $U$ as $\propto U^{-14}$. We use the lowest $N=50$ eigenvalues in the $S^z = 2$ sector for the optimization with the learning rate $\alpha = 0.1$. }
       \label{HUBBARD HIGHER ORDER}
  \end{figure}

In Fig.~\ref{HUBBARD HIGHER ORDER} we show the $U$ dependence of the difference between the optimized parameters and the prediction from the perturbation theory, $|J_i -J^{ p}_i|_{i=1,2,3}$, $|K_i -K^{p}_i|_{i=2,3}$ and the cost evaluated for the cross-validation set. We see that at large $U$, the differences of the couplings decay as $U^{-7}$ and the cross-validation error drops as $U^{-14}$. The $U^{-14}$ dependence of the cost, or equivalently the $U^{-7}$ deviation in the spectrum indicates that the perturbative spin model presented in Ref.~\onlinecite{PhysRevB.37.9753} is the right one and our ML scheme can reproduce it properly. 

Here we comment on the finite size effect. From the perturbation theory viewpoint, the effective spin interactions originate from the virtual hopping processes in the original Hubbard model. Thus, when the system size is $L$, finite size effects can only appear at $O(t^L/U^{L+1})$ if we impose the periodic boundary condition. This nature allows us to construct the effective spin model quite accurately even using such relatively small system size as we have seen above. 

\subsection{Periodically modulated Hubbard model}\label{3-b}
Concerning the homogeneous Hubbard chain, we show that the ML scheme works effectively and we could resolve the discrepancy among high order perturbative calculations in the previous works. Next, we turn on the inhomogeneity in Eq.~\eqref{Hubbard}, which breaks the translational symmetry. For simplicity, we take a large repulsive interaction where only the nearest-neighbor spin interaction would matter and we can use the perturbative results as a reference. We take the trial spin Hamiltonian in the form:
\begin{align}
H = E_c + \sum_{i=1}^{L} J_{1,i} \bm{S}_{i}\cdot \bm{S}_{i+1}.\label{EFF SPIN}
\end{align}
The exchange couplings are optimized in a site-dependent way so that the number of parameters scales with the system volume (precisely given by $L+1$ including $E_c$). We take the following initial values inferred from the second order perturbation theory of the homogeneous Hubbard model: $J_{1,i} = 4(1+0.01\xi_i)t^2/U_i$, where $\xi_i$ is a random number taken from the standard normal distribution $\mathcal{N}(0,1)$. The randomness is introduced to resolve the ambiguity of the phase of the spatial modulation in the spin model (we only use the energy spectrum of a periodic Hubbard chain). The initial value for the offset $E_c$ is taken to be in the middle of the energy window $[E_0, E_{N-1}]$, $E_c = (E_0 + E_{N-1})/2$.  Here we are treating $E_c$ as an optimization parameter to show that the way we treat the constant shift $E_c$ (fixing by equating the averages or treating as a parameter to be optimized) is not essential.

\begin{figure}[htbp]
   \centering
\includegraphics[width = 90mm]{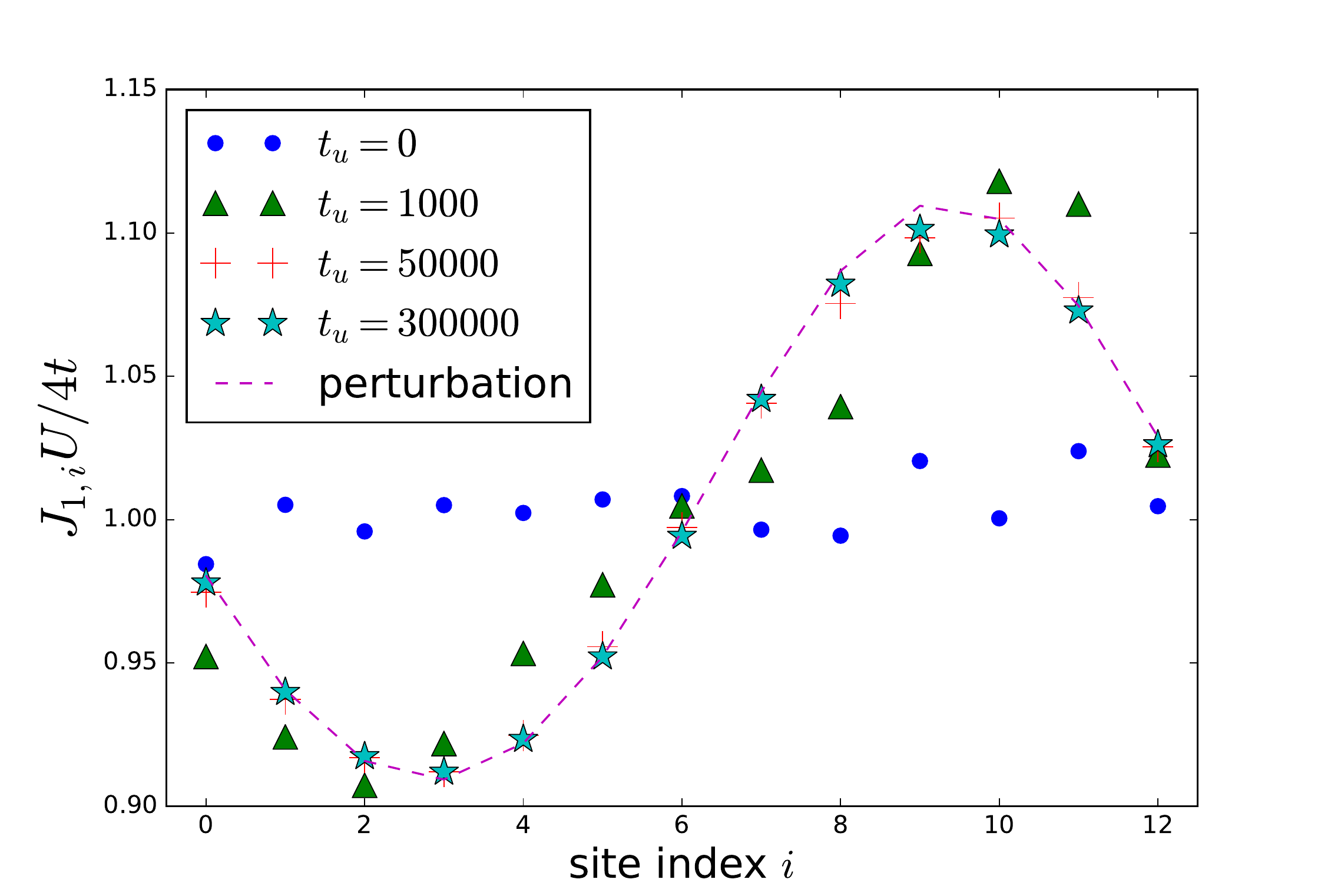}
       \caption{Spatial profile of the nearest-neighbor exchange couplings estimated from the lowest $N=30$ eigenvalues ($S^z = 4.5$ sector) of the Hubbard model Eq.~\eqref{Hubbard} with $U_n =  50\left[1 -0.1 \sin (2\pi n/L)\right]$ ($n = 1, ..., L = 13$). We use $\alpha = 0.3$ and update the parameters $t_{u} = 300,000$ times and compare the estimated parameters (markers) with those expected from the second order perturbation theory (broken line). The optimized parameters reproduce the spatial modulation of the original Hubbard model.}
       \label{HUBBARD BENCHMARK}
  \end{figure}

We consider the Hubbard chain of length $L= 13$ with the repulsive interaction $U_n = U [1-0.1\sin (2\pi n/L)]$, where $U/t=50$. We take the learning rate $\alpha = 0.3$ and update the parameters $J_{1, i}$ using the proposed update scheme Eq.~\eqref{algorithm} for the lowest $N =  30$ eigenvalues in the $S^z = 4.5$ sector.
In Fig.~\ref{HUBBARD BENCHMARK}, we present the spatial profile of the estimated exchange couplings $J_{1,i}$ for $i = 1, ..., L$. We see that as the learning proceeds, the estimated exchange couplings approach to $J_{1,i}= 4t^2/U_i$~\footnote{As we mentioned, the phase of the spatial modulation of the spin model is ambiguous. In Fig.~\ref{HUBBARD BENCHMARK} we choose the phase $\phi$ as a fitting parameter and plot $J_{1,i} = 4t^2/U_i$ with $U_n = U [1-0.1\sin (2\pi n/L + \phi)]$ for the $\phi$ reproducing the $t_u = 300000$ result best.}, which we naively expect from the perturbation theory of the homogeneous Hubbard model. It demonstrates that our scheme works even in the absence of the translational symmetry and can extract the information of the symmetry breaking only from the low-lying spectrum. It also implies that our method works for more generic boundary conditions without translational invariance.

\section{Construction of entanglement Hamiltonian} \label{HA construction} 
In this section, we show that our approach works for the construction of the EH from the ES of a given quantum many-body state.

First, we briefly review the definition of the ES and the EH. Let us consider a spin system consisting of two subregions $A$ and $B$.

For a given quantum  state $\ket{\Psi}$ defined on $A\cup B$, we introduce the reduced density matrix of the subregion $A$ by tracing out the spin degrees of freedom in the region $B$:
\begin{align}
\rho_A &= {\rm{Tr}_{B}} \ket{\Psi}\bra{\Psi}. 
 \label{RDM}
\end{align}
The EH is defined as the Hamiltonian whose density matrix at the unit temperature is the reduced density matrix:  $\rho_A \equiv e^{-H_A}$. The ES is defined as its eigenvalues~\cite{PhysRevLett.101.010504} and reflects the non-local quantum correlation between the subregions $A$ and $B$.

The ES and EH are defined through the entanglement cut, the {\it virtual} separation of the given quantum state into two parts. Nevertheless, there is an intriguing
conjecture that they are related to physical edge spectrum and
Hamiltonian which appear when the system is actually (not virtually)
cut into two parts~\cite{PhysRevLett.101.010504,PhysRevLett.108.196402}.
This conjecture is confirmed in,
for example, the two-dimensional topological phase, where there appear chiral edge modes described by the conformal field theory at the physical boundary. In lattice systems, the canonical example of this correspondence is found in the Affleck-Kennedy-Lieb-Tasaki (AKLT) chain~\cite{PhysRevLett.59.799,Affleck:1988aa,PhysRevB.81.064439,PhysRevB.83.075102,PhysRevB.85.075125}, a variant of the $S = 1$ antiferromagnetic Heisenberg chain. The ground state of the AKLT chain is written in terms of singlet bonds between the emergent $S = 1/2$ degrees of freedom, and on both the virtual and the physical edges, there appear the free $S = 1/2$ degrees of freedom.

As we mentioned in the introductory part, EH cannot be uniquely determined from
a given ES, even in principle.
However, as implied by the entanglement/edge correspondence
conjecture~\cite{PhysRevLett.101.010504,PhysRevLett.108.196402},
we may expect EH to be ``local'', as long as the original model is so.
With the requirement of the locality, EH can be practically
estimated by the ES, as we will demonstrate
below.

\subsection{SU(2) symmetric Heisenberg ladder}
\begin{figure}[htbp]
   \centering
\includegraphics[width = 80mm]{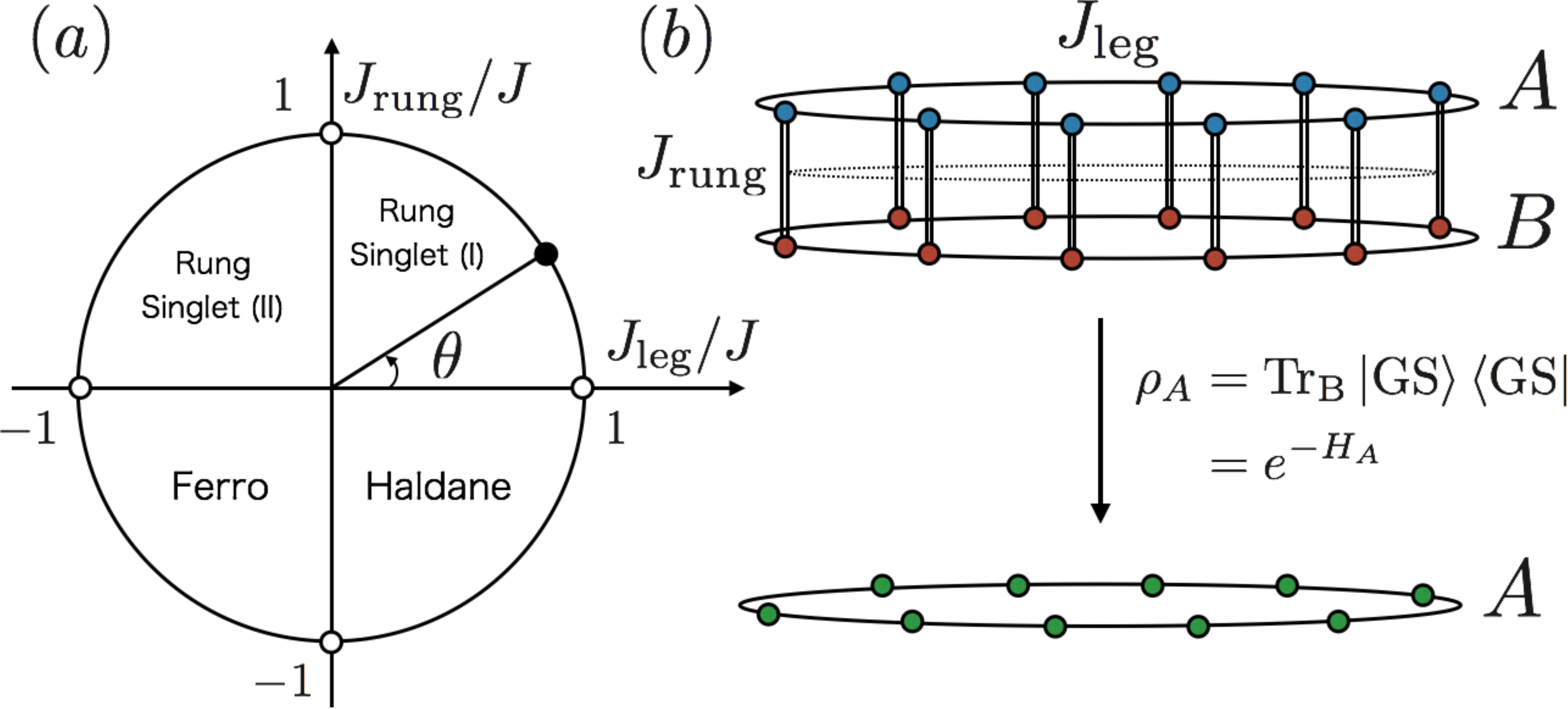}
       \caption{($a$) Ground state phase diagram of the model Eq.~\eqref{HA}. ($b$) Schematic diagram of the definition of the entanglement Hamiltonian for the leg-separating cut. The dotted line indicates the entanglement cut.}
       \label{Two-leg-ladder}
  \end{figure}

As a concrete example, let us discuss
the two-leg periodic $S = 1/2$ Heisenberg ladder
with $A$ and $B$ legs
\begin{align}
H &= \sum_{i=1}^{L}  J_{\rm rung}\bm{S}_{i, A}\cdot\bm{S}_{i, B} \nonumber \\ &+\sum_{i=1}^{L}\sum_{j= A, B}J_{\rm leg}\bm{S}_{i, j}\cdot\bm{S}_{i+1, j}.  \label{Ladder}
\end{align}
We introduce two kinds of interactions, $J_{\rm leg}$ and $J_{\rm
rung}$. The former describes the intra-chain interaction and the latter
does the inter-chain one. The ground state phase diagram of
Eq.~\eqref{Ladder} for $J_{\rm leg} = J \cos\theta$, $J_{\rm rung} = J
\sin\theta$ is given in Fig.~\ref{Two-leg-ladder}($a$)~\cite{Barnes:1993aa,Dagotto:1996aa,PhysRevLett.105.077202}.

Following Refs.~\cite{PhysRevB.83.085112,PhysRevLett.105.077202,PhysRevB.85.054403,PhysRevB.88.245137,PhysRevB.83.245134},
here we study the entanglement for a virtual cut which separates the chains
as shown in Fig.~\ref{Two-leg-ladder} ($b$).
The EH then acts on the Hilbert space of a single $S=1/2$ chain.
In Ref.~\cite{PhysRevLett.105.077202}, it was argued that
the EH is qualitatively given by the $S=1/2$ Heisenberg chain based on the numerically obtained ES.
However, since the identification was made by a visual inspection
of the spectrum, it is difficult to analyze its accuracy and possible corrections to the simple Heisenberg model quantitatively. If we have access to all the eigenvectors of the reduced density matrix, as was performed in Ref.~\onlinecite{PhysRevB.83.245134} we can construct the EH exactly within a particular ansatz. However, the full-diagonalization of large matrices is computationally heavy and its applicability is limited to small subsystems. In the following, we show that our ML scheme offers another approach for the construction of the EH. In the final part of this section, we compare the ML approach with the full-diagonalization approach in detail. 

Below we focus on the ground state of the model Eq.~\eqref{Ladder} and estimate the EH using the $N$ smallest ES, not the entire eigenvalues and eigenvectors of the reduced density matrix. As the trial Hamiltonian for $H_A$, we use the following $SU(2)$ symmetric spin Hamiltonian,
\begin{align} \label{HA}
H &=E_c+ \sum_{i=1}^{L}\sum_{\delta=1,...,5} J_{\delta} \bm{S}_i \cdot \bm{S}_{i+\delta} \\ 
&+\sum_{i=1}^{L}\sum_{\alpha,\beta,\gamma = \{1,2,3\}}\frac{K_\alpha}{2}(\bm{S}_i\cdot \bm{S}_{i+\alpha})(\bm{S}_{i+\beta} \cdot \bm{S}_{i+\gamma}) \nonumber 
\end{align}
where $E_c$, $J_{\delta}$, and $K_\alpha$ are the parameters to be determined. In the following, we fix $E_c$ by equating the average of the input eigenvalues to that of the ansatz.  Here the summation over $\alpha, \beta, \gamma = 1,2,3$ is taken under the condition $\alpha \neq \beta \neq \gamma$ and $\alpha \neq \gamma$. We take into account all the Heisenberg interactions (in a system of $L = 10$) and the three simplest four-spin interactions. In the previous work~\cite{PhysRevB.83.245134} based on the full-diagonalization of the reduced density matrix, the two-body interactions are fully incorporated, but multi-spin interactions were treated all together and their individual contributions were not investigated. Below, using the ML, we show the four-spin interactions are important for understanding the ES.
 
When the rung interaction is dominantly strong and antiferromagnetic ($\theta \simeq \pi/2$), for each $i$, the spins in the two legs $\bm{S}_{i, A}$ and $\bm{S}_{i,B}$ form a singlet, so the virtual cut produces $S=1/2$ spins at each singlet bond just like in the AKLT model. In the presence of the strong rung repulsion, therefore, we have well-defined $S=1/2$ degrees of freedom at each site in the EH. Away from $\theta = \pi / 2$, tracing out the leg $B$ makes the EH of the leg $A$ non-local. In particular, the situation is quite nontrivial when the rung interaction is weakly antiferromagnetic or ferromagnetic. In these cases, long-range and multi-spin interactions would be important to reproduce the ES. In other words, the ansatz Eq.~\eqref{HA} becomes no longer the proper one and the cost becomes large.

The estimation of the EH proceeds as follows. We first obtain the ground state wave function of the model Eq.~\eqref{Ladder} in the $S^z_{\rm tot} = \sum_i S^z_i = 0$ sector and calculate the reduced density matrix $\rho_A$ by tracing out the chain $B$. The ES is obtained by diagonalizing $\rho_A$ and we optimize the trial spin Hamiltonian Eq.~\eqref{HA} using Eq.~\eqref{algorithm}. We consider the ladder of length $L = 10$, and take the $N = 100$ smallest eigenvalues in  the three magnetization sectors $S^z_{\rm A} = \sum_{i \in A}S^z_i = 0, 1, 2$ for the learning~\footnote{The entanglement cut considered here preserves the $z$ component of the total spin, so that the EH can be block-diagonalized in terms of  $S^z_{\rm A}$. The choice of the magnetization sector (i.e. which $S^z_{\rm A}$ sector) is arbitrary and one can combine several sectors to avoid the trap to the local minima as we do here.}. We use the stochastic gradient descent method with {\it Adam}~\cite{Adam} under the following hyperparameters $\alpha = 0.001, \beta_1 = 0.99, \beta = 0.999, \epsilon = 10^{-8}$ (see appendix B).

 In Fig.~\ref{ES_ESTIMATE} we show the $\theta$ dependence of the cross-validation error Eq.~\eqref{Cost_function} evaluated for the $N=40$ eigenvalues in the $S^z_A = 3$ sector (a sector not used for the learning), and the estimated parameters. To compensate the difference in the energy scale of the ES, we normalize the cross-validation error with the width of the energy window squared $(E_{39} - E_0)^2$. 
 
 When both the rung and the leg interactions are ferromagnetic ($-\pi < \theta <-\pi/2$) the ground state is
ferromagnetically ordered. In the following, we discuss the results in the other regions of the phase diagram Fig.~\ref{Two-leg-ladder}($a$).

 \subsubsection{Antiferromagnetic rung region, $0 < \theta  < \pi$}
 
When $\theta = \pi/2$, the system turns into decoupled $L$ singlets
and the ES is perfectly degenerate so that the EH is of the free $S=1/2$ spins. In Fig.~\ref{ES_ESTIMATE}, we see that even away
from this trivial limit, the short-ranged ansatz Eq.~\eqref{HA} well reproduces the ES when the rung is antiferromagnetic.
In the limit $\theta \to \pi/2$, we find that the EH is dominated by the nearest-neighbor Heisenberg
interaction $J_1$ (all the other couplings in Eq.~\eqref{HA} become vanishingly small as compared to $J_1$) so that the entanglement/edge correspondence is realized. We note that the sign of Heisenberg couplings shows no frustration in consistent with Refs.~\onlinecite{PhysRevLett.105.077202,PhysRevB.83.245134}.

The cost grows as $\theta$ approaches to $\theta = 0, \pi$, and at the same
time, coupling constants of further neighbor and multi-spin interactions
become large. Near $\theta = \pi/2$ the largest correction to the nearest-neighbor Heisenberg model comes from the second neighbor
Heisenberg interaction in consistent with the previous study~\cite{PhysRevB.85.054403} using the perturbation theory near the strong rung limit and the study taking the full-diagonalization approach~\cite{PhysRevB.83.245134}. However, as $\theta$ approaches to zero, we see that the four-spin interaction terms such as $K_1$ and $K_3$ grow rapidly and become comparable or even larger than the second-neighbor Heisenberg interaction.
Such large contributions from the four-spin interaction terms have not been known in the previous works. The large error and the growth of coupling constants imply that, far away from $\theta \simeq \pi/2$, the short-ranged Hamiltonian Eq.~\eqref{HA} is no longer a valid ansatz.

 \subsubsection{Ferromagnetic rung region, $-\pi/2 < \theta  < \pi$}
This region is called the Haldane phase, since the ladder with $\theta \simeq -\pi/2$ is approximately described as the antiferromagnetically coupled $S=1$ chain, whose ground state is known to be in the Haldane phase. In the earlier studies~\cite{PhysRevLett.105.077202,PhysRevB.85.054403,PhysRevB.83.245134},
it was argued that the EH in this phase is also similar to
the antiferromagnetic Heisenberg chain.

Figure ~\ref{ES_ESTIMATE} shows that the cross-validation error under the short-ranged 
Hamiltonian ansatz Eq.~\eqref{HA} is large for all $\theta$ in this
region $-\pi/2 < \theta <0$. Moreover, the estimated coupling constants
show non-systematic $\theta$ dependence, contrary to the antiferromagnetic rung case. The second neighbor Heisenberg coupling is estimated to be very small, but this is inconsistent with the exact diagonalization study~\cite{PhysRevB.83.245134}. That is, the ML scheme under the short-ranged EH ansatz fails to construct the EH in this region.

In the previous study~\cite{PhysRevB.83.245134}, it was observed that the Heisenberg interaction is long-ranged as there is a non-negligible contribution from the 7th neighbor Heisenberg interaction, which is the longest one in their $L = 14$ system, while the authors claimed that the multi-spin interactions would be small. Here our results support the highly non-local nature of the EH in this phase and demonstrate that the multi-spin interactions are in fact not at all small compared to the two-body interactions. The EH in the Haldane phase is, therefore, highly non-local and has contributions of large multi-body interactions. To discuss the properties of the EH, we have to incorporate many other terms such as further neighbor four-body interactions, six-body interactions, and so on, so discussing the entanglement/edge correspondence based on the simple model only with two-body interactions~\cite{PhysRevLett.105.077202,PhysRevB.85.054403,PhysRevB.83.245134} is insufficient.

\subsection{Anisotropic Heisenberg ladder: restoration of the entanglement/edge correspondence}

\begin{figure}[htbp]
   \centering
\includegraphics[width = 90mm]{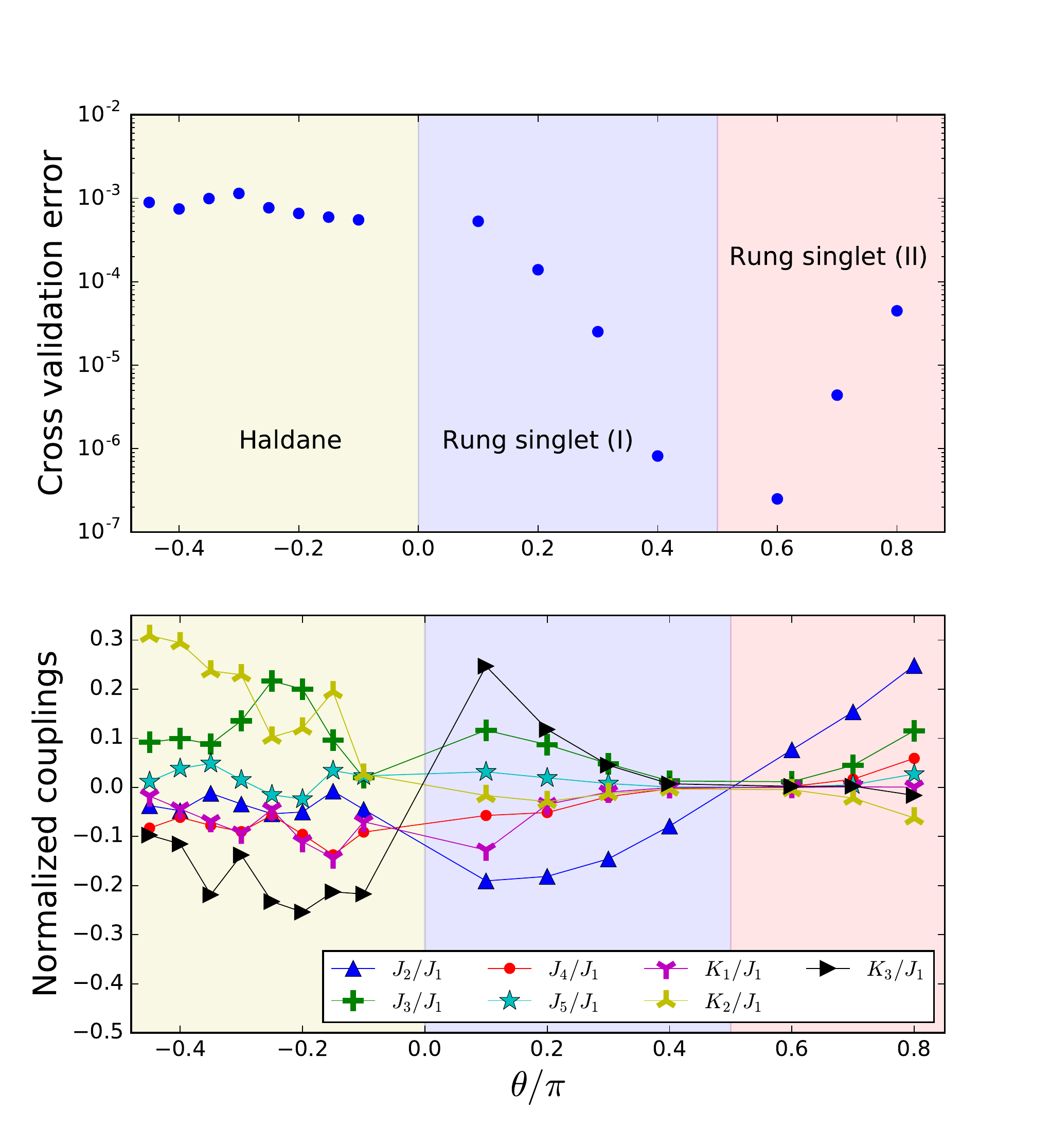}
       \caption{ Cross-validation error calculated with the lowest $N = 40$ eigenvalues in the $S_A^z = 3$ sector and the estimated couplings $J_\delta/J_1$ ($\delta = 2,...,5$), $K_\alpha/J_1$ ($\alpha = 1,2,3$) as functions of $\theta$. The cross-validation error is normalized with the square of the energy window of the entanglement spectrum. Near $\theta = \pi/2$, the entanglement spectrum is almost flat, and the nearest-neighbor Heisenberg model well reproduces the spectrum. As $\theta$ deviates from this special point, long-range and many-body interactions become important and the cross-validation error increases.}
       \label{ES_ESTIMATE}
  \end{figure}

What is particularly notable in Fig.~\ref{ES_ESTIMATE} is that the non-local or multi-spin interactions still have large values even near $\theta = -\pi/2$, where the ground state is represented as a collection of independent spin triplets. This is in significant contrast to the $\theta = \pi/2$ case, where the ES is flat and all the coupling constants in the EH vanish. The non-locality of the EH (or the dispersive ES) in this region is, therefore, not originating from the leg-interaction. As we show in appendix C, the origin of the dispersive ES in this limit is the superposition of spin triplet states in the ground state wave function of the Haldane phase. The EH in the Haldane phase is, therefore, intrinsically non-local and has less physical meaning than that in the rung singlet phases in the context of the entanglement/edge correspondence. There we cannot expect that the EH is short-ranged and our ML approach works.

Indeed, by breaking the SU(2) symmetry and turns the ground state into the large-$D$ phase, where the state is smoothly connected to the product state in the leg direction, we can make the EH local and restore the entanglement/edge correspondence. To show that, let us introduce the XXZ-type anisotropy in the rung interaction:

\begin{align}
H &= \sum_{i=1}^{L}  J_{\rm rung}(S^x_{i, A}S^x_{i, B}+S^y_{i, A}S^y_{i, B}+ (1 + D) S^z_{i, A}S^z_{i, B}) \nonumber \\ &+\sum_{i=1}^{L}\sum_{j= A, B}J_{\rm leg}\bm{S}_{i, j}\cdot\bm{S}_{i+1, j}.  \label{Ladder_anis}
\end{align}

For $\theta \simeq -\pi/2$ and $(1 + D) < 1$, the state with $(S^z_{i, A} + S^z_{i, B}) = 0$ for each $i$ is favored and the ground state is approximately given by the product of triplet $S^z=0$ state $\ket{\Psi}_i = (\ket{\uparrow_A \downarrow_B}_i + \ket{\downarrow_A \uparrow_B}_i)/\sqrt{2}$ of each site $i$. Then the reduced density matrix becomes almost proportional to the identity matrix and the ES is almost flat, similar to the case of $\theta \simeq \pi/2$. The flat ES indicates the locality of the EH, and we can apply the ML scheme to construct the EH by adapting the short-ranged EH ansatz. Below we take $\theta = -0.48\pi$ and investigate the $D$ dependence of the cross-validation error and the estimated couplings. For $\theta  = -0.48 \pi$, the ground state is in the large-$D$ phase for $D \lesssim -0.05$~\footnote{We determine this critical value $D_c \sim -0.05$ for $\theta=-0.48\pi$ by using the level spectroscopy~\cite{0305-4470-28-19-003,1742-6596-320-1-012018}} For simplicity, we consider the following ansatz for the EH, which contains XXZ-type couplings up to fifth neighbors (We take $L$ = 10): 

\begin{align}
H &= \sum_{k = 1, ..., 5}\sum_{i=1}^{L}  J_{k}\bm{S}_{i}\cdot\bm{S}_{i+k}+\Delta_k S^z_{i}S^z_{i+k}\label{Ansatz_wo_SU2},
\end{align}

\begin{figure}[htbp]
   \centering
\includegraphics[width = 80mm]{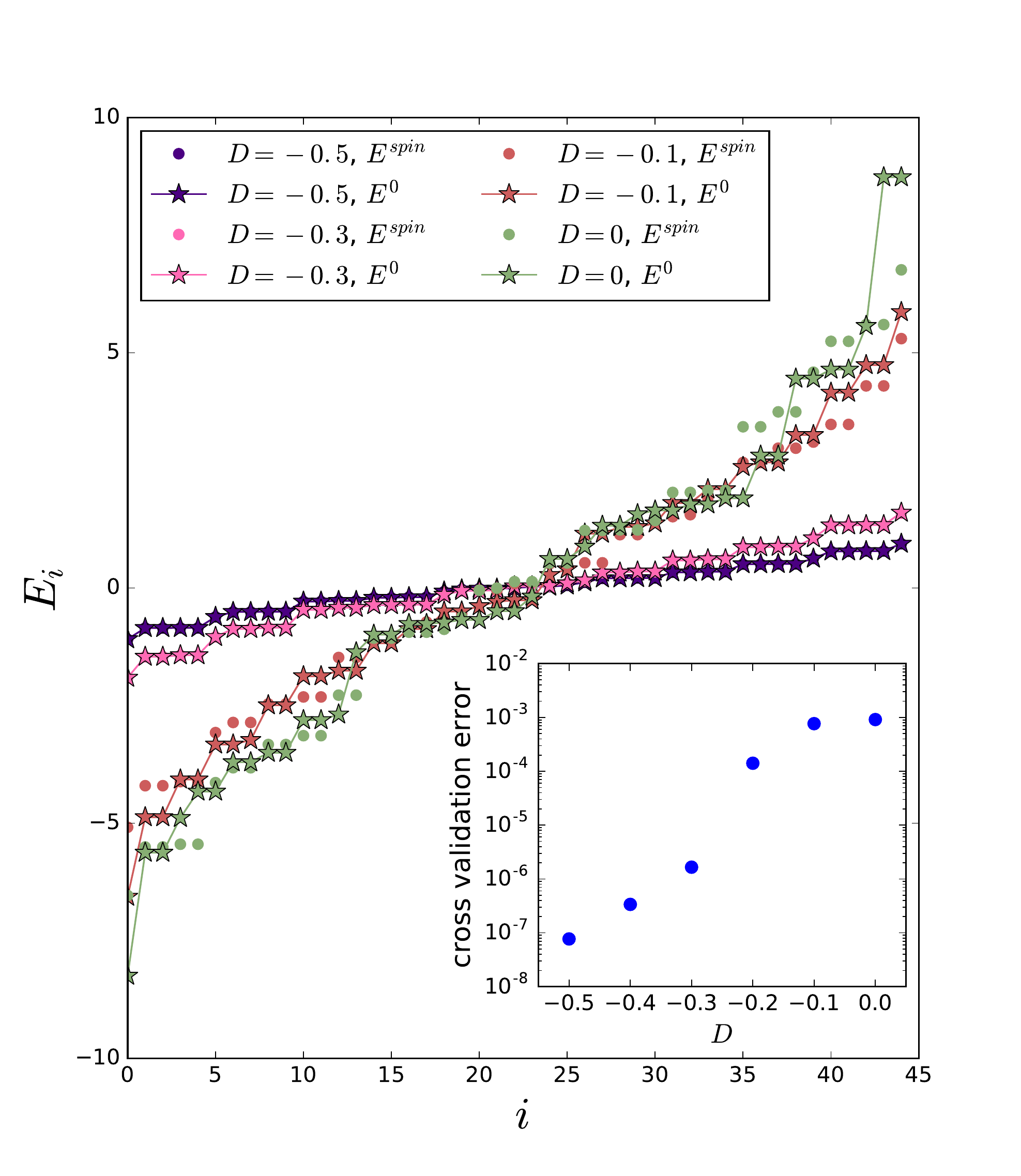}
       \caption{Estimation of the entanglement Hamiltonian of the two-leg spin ladder with length $L=10$ in the presence of the XXZ-type anisotropy $D$. As input data for the learning, we use 100 eigenvalues from $S^z_A = 0, 1, 2$ sectors respectively and optimized the parameters in the ansatz Eq.~\eqref{Ansatz_wo_SU2}, using Adam. As we increase the anisotropy (goes deep inside the large-$D$ phase), the entanglement spectrum $\{E^0_i \}$ (in the $S^z_A = 3$ sector) becomes flatter and the energy spectrum of the estimated spin model $\{E^{spin}_i\}$ approaches to that, as is quantitatively confirmed with the cross-validation error (inset) calculated in the same way as Fig.~\ref{ES_ESTIMATE}. We note that the cross-validation error is normalized with the square of the energy window of the entanglement spectrum $(E_{39} - E_0)^2$ in the $S^z_A = 3$ sector.}
       \label{ES_anis}
  \end{figure}

Just as in Fig.~\ref{ES_ESTIMATE}, here we use $N=100$ eigenvalues in the $S^z_A = 0,1,2$ sectors for the parameter update and use the $N=40$ eigenvalues in the $S^z_A = 3$ sector for the cross-validation. Figure~\ref{ES_anis} shows that as we decrease $D$ from zero, the ES becomes less dispersive and the cross-validation error decreases significantly. Namely, the EH becomes more and more local and the one constructed with the ML reproduces the true EH better, as the ground state goes deep inside the large-$D$ phase. We note that the cross-validation error is normalized with the square of the width of the energy window $E_{39} - E_0$, so that the improvement of the cross-validation error is not from the change in the energy scale of the ES. The dominant term in the estimated EH is the nearest-neighbor Heisenberg interaction. Given the low cross-validation error for the simple ansatz Eq.~\eqref{Ansatz_wo_SU2}, we conclude that the anisotropy restores the locality in the EH and the entanglement/edge correspondence.

In the framework of the field theory (bosonization)~\cite{PhysRevB.88.245137}, the Rung Singlet (I) and Haldane phases of the Heisenberg ladder (without the anisotropy) are almost equivalent, since they differ only in the location of the potential minimum to which the boson field is pinned. The ES in these phases is shown to be given by the energy spectrum of a Tomonaga-Luttinger liquid. As we have discussed above, the dispersive ES in the Haldane phase has the physical origin different from the one in the Rung Singlet (I) phase. Although the apparent long-range and multi-spin interactions in the EH we obtain for the Haldane phase do not contradict this field-theoretical correspondence because the field theory only focuses on the low-energy limit, it is still nontrivial how these two pictures hold altogether. In other words, there would be still many uncovered issues related to the entanglement/edge correspondence beyond the low-energy, or field-theoretical region, where our ML approach could be a powerful tool for their studies.

In the above, we consider the simplest entanglement cut, which preserves the translational symmetry in the leg direction. However, since the choice of the boundary condition is arbitrary in the present scheme, we can take arbitrary entanglement cut, not being translationally invariant in general.
\subsection{Discussion}

Finally, we compare our ML approach for the construction of the EH with the full-diagonalization approach~\cite{PhysRevB.83.245134} in more detail. When the subsystem size is small and all the eigenvectors of the reduced density matrix are available, the full-diagonalization approach gives a more reliable estimate for the EH, since it is, in principle, exact. However, as we mentioned, the difficulty comes into play when one tries to estimate the EH of large subsystems. In particular, when the translational symmetry is absent, which is the generic situation due to the entanglement cuts, the possible subsystem size is at present up to about 21-22 sites of $S = 1/2$ spins even if we assume the conservation of the total magnetization. By contrast, our ML scheme only uses the low-lying spectrum of the reduced density matrix, so that one can exploit the computationally cheap algorithms for the matrix diagonalization such as the Lanczos method and construct the EH for much larger subsystems. Hence, the full-diagonalization approach~\cite{PhysRevB.83.245134} and our ML approach will work complementary for the reliable estimation of the EH. That is, the former is particularly suitable for small subsystems and can determine the reasonable form of the ansatz of the EH, while the latter can optimize that ansatz for much larger subsystems. 

It is also worth mentioning that in the presence of the symmetries like the conservation of the subsystem magnetization $S^z_A$, choosing the proper sector of that symmetry further increases the possible subsystem size in the ML approach. Our approach can be applied to the low-lying spectrum of a particular symmetry sector or a combination of them, so that we can avoid dealing with the entire subsystem with a large Hilbert space dimension. This could change the bottleneck of the construction of the EH from the diagonalization of the reduced density matrix to the calculation of the (ground state) wave function. The wave function of the ground state (and also the low-lying states) of many-body systems is sometimes available for quite large system sizes by using the modern algorithms such as the Density Matrix Renormalization Group (DMRG)~\cite{PhysRevLett.69.2863,RevModPhys.77.259}. Therefore, combined with the DMRG, our ML approach could give the EH of the extremely large subsystems. 

The flexibility to the boundary condition and the accessibility to large systems of the ML scheme also allow us to study the EH in higher dimensions with changing the geometry of the subsystem. For example, it is known that the entanglement entropy of a (2+1)-dimensional conformal field theory acquires an additional universal logarithmic contribution~\cite{PhysRevLett.97.050404,CASINI2007183} if the subregion in which the entanglement is calculated has sharp edges. It is an interesting future study to investigate how the universal corner entanglement entropy in the field theory is implemented in the lattice EH, for which the ML approach could play an important role. 

\section{Conclusion and outlook}
In this paper, we developed a machine learning algorithm to construct (spin) Hamiltonians from the energy or entanglement spectra. As a benchmark, we first consider the Hubbard models and use the proposed scheme to estimate the effective spin Hamiltonians. For the homogeneous Hubbard model, we show that the machine learning gives an effective Hamiltonian consistent with the sixth order perturbation theory presented in Ref.~\onlinecite{PhysRevB.37.9753}. Although the sixth order terms have been inconsistent among papers, on the basis of the quantitative error analysis, we show that one in Ref.~\onlinecite{PhysRevB.37.9753} is the correct one. We also demonstrate that from the energy spectrum we can infer the breaking of the translational symmetry in the periodically modulated Hubbard model. 

Moreover, we show that the same algorithm can be applied to construct the entanglement Hamiltonian from the entanglement spectrum of a given quantum many-body state, taking the two-leg antiferromagnetic Heisenberg ladder as the example. We observe a qualitative difference between the Rung Singlet (I) and Haldane phases, which goes beyond the field-theoretical description. We associate the observed difference to the different origins of the entanglement in these phases. Since the machine learning approach only uses the low-lying entanglement spectrum and does not require the full-diagonalization of the reduced density matrix, it can be applied to large subsystem sizes, in particular in the presence of symmetries. If combined with the Density Matrix Renormalization Group method, it offers a way to examine the celebrated entanglement/edge correspondence for the unprecedentedly large system sizes.

So far, we discussed the ML construction from the numerically obtained spectrum. However, the applicability of the ML approach is not limited to that, and we can consider the use for experimental data. Throughout this paper, we use the mean squared error of the energy spectrum as the cost function and optimize the Hamiltonian. However, the choice of the cost function or the quantity to be optimized is not restricted to that. The heart of the presented strategy is the fact that the gradient of the energy spectrum can be analytically obtained using the knowledge of the perturbation theory in quantum mechanics. Therefore, as long as the cost is defined as a function of the energy eigenvalues, essentially the same procedure gives the optimized Hamiltonian. For example, we could use the density of states (or specific heat) to define the cost function and optimize the effective Hamiltonian.

Another important application of our method is for the study of the strongly-correlated electron systems. Although we focused on the one-dimensional systems in Sec.~3 for simplicity, our approach should work for general dimensions. The Hubbard model in two dimensions is one of the most important subjects in the modern condensed matter physics in the context of high-temperature superconductivity in cuprates~\cite{RevModPhys.66.763} and quantum spin liquids~\cite{PhysRevLett.91.107001,PhysRevLett.100.136402,PhysRevB.79.235130,PhysRevB.79.235130,PhysRevLett.105.267204,Sorella:2012aa,Han:2012aa}.  The main advantage of our approach is that it can be readily applied to the $t/U = O(1)$ region where various nontrivial physics arise from the competition between the kinetic energy and the interaction energy. In this region, perturbative treatment breaks down or one has to calculate quite high order terms. On the other hand, our scheme works as long as a spin model describes the low-energy physics of the Hubbard model. Even when the original Hubbard model is more complicated, e.g. contains the further neighbor hoppings, long-range interactions, or disorders, the present method can be straightforwardly applied to those deformed models. 

At the half-filling, the dimension of the Hilbert space of the Hubbard chain with length $L$ is reduced from $4^L$ to $2^L$ by modeling the low-energy physics by the spin model. Therefore, the derivation of the effective spin model can be regarded as a compression of the Hilbert space and doubles the achievable system size in numerical calculations of the Hubbard-like models. As a result, with our numerical optimization approach, one can investigate the low-energy properties of the Hubbard model at the half-filling with unconventionally large system sizes. Although the size dependence of the optimized parameters obtained by our method may require careful treatments, this would offer a new approach for the various unsolved issues in the strongly-correlated electron systems, such as the ground state property of the Hubbard-like systems on the triangular lattice~\cite{PhysRevLett.91.107001,PhysRevLett.99.127004,PhysRevLett.100.136402,PhysRevLett.105.267204,PhysRevB.92.140403}.

Finally, we comment on the inverse problem of the construction of the low-energy model: estimation of the parent Hamiltonian for a given low-energy model. In the context of spin liquids, for example, it is common that we only have a low-energy model written in terms of the emergent degrees of freedom different from electrons or spins such as dimers, slave bosons, or lattice gauge fields~\cite{PhysRevB.35.8865,PhysRevB.65.165113}. Although we focused on the construction of the effective model from the parent Hamiltonian in this paper, our method can be also applied to the estimation of the parent Hamiltonian. Since it is of fundamental importance to have the Hamiltonian written in terms of the physical degrees of freedom such as electrons or spins for the material search, our results could help experimentalists materialize various novel quantum states.

\section{Acknowledgement}
We thank M. Ohzeki, K. Okamoto, S. Furukawa, Y. Fuji, D. Poilblanc, and M. Takahashi for useful comments and discussions. H. F. and Y. O. N are supported by Advanced Leading Graduate Course for Photon Science (ALPS) of Japan Society for the Promotion of Science (JSPS). The works of H. F., Y. O. N., and S. S are supported by JSPS KAKENHI Grant-in-Aid for JSPS Fellows Grant No.~JP16J04752, No.~JP16J01135, and No.~JP15J11250, respectively. The work of M. O. is supported in part by JSPS KAKENHI Grant No. 16K05469.
A part of the computation in this work has been done using the
facilities of the Supercomputer Center, the Institute for Solid State
Physics, the University of Tokyo.

\widetext{
\appendix

 \section{Sparse modeling approach}
 In this appendix, we explain how to practically choose the ansatz to be optimized by taking the Hubbard model as an example.  As is discussed in the main text, it is mathematically impossible to determine the effective model from nowhere if one only use the spectrum. In other words, even if we fix the Hilbert space, there are still an infinite number of Hamiltonians which give exactly the same spectrum. Therefore, setting an ansatz is essentially important.
 
 Let us consider the one-dimIn ensional Hubbard model and take the following ansatz:
 \begin{align}
H_{\rm spin} &= E_c +\sum_{i=1}^{L}\sum_{\delta=1,2,3} J_{\delta} \bm{S}_i \cdot \bm{S}_{i+\delta} 
+\sum_{i=1}^{L}\sum_{\alpha,\beta,\gamma = \{1,2,3\}}\frac{K_\alpha}{2}(\bm{S}_i\cdot \bm{S}_{i+\alpha})(\bm{S}_{i+\beta} \cdot \bm{S}_{i+\gamma}) \nonumber \\
&+\sum_{i=1}^{L}M_1(\bm{S}_i\cdot \bm{S}_{i+1})(\bm{S}_{i+3} \cdot \bm{S}_{i+4}) +M_2(\bm{S}_i\cdot \bm{S}_{i+1})(\bm{S}_{i+2} \cdot \bm{S}_{i+4}) +
M_3(\bm{S}_i\cdot \bm{S}_{i+2})(\bm{S}_{i+3} \cdot \bm{S}_{i+4})\nonumber \\ &+\sum_{i=1}^{L}M_4(\bm{S}_i\cdot \bm{S}_{i+2})(\bm{S}_{i+1} \cdot \bm{S}_{i+4})+
M_5(\bm{S}_i\cdot \bm{S}_{i+3})(\bm{S}_{i+2} \cdot \bm{S}_{i+4}) + M_6(\bm{S}_i\cdot \bm{S}_{i+3})(\bm{S}_{i+1} \cdot \bm{S}_{i+4})
\label{REG_HAM}
\end{align}
  Here the summation over $\alpha, \beta, \gamma = 1,2,3$ is taken under the condition $\alpha \neq \beta \neq \gamma$ and $\alpha \neq \gamma$. This ansatz is fairly complex. It contains Heisenberg interactions up to third neighbors and the nine different four-spin interactions. As is trivial, increasing the number of parameters always decreases the cost  evaluated for the given dataset. However, this leads to the overfitting and diminishes the generalization ability of the obtained model. In our context, the generalization ability means how well the obtained model describes the low-energy physics of the original model, not only the given low-energy spectrum. In order to avoid the overfitting, it is effective to pick up ``important terms" to explain the given data. For that purpose, here we consider the $L_1$ norm regularization of the cost function:
\begin{equation}
{\rm RegCost }(\{ c_j\}) =  \frac{1}{2N}\sum_{i = 0}^{N-1} \left[E^{\rm data}_i  - E_i (\{c_j \})\right]^2  +\lambda\sum_{j}{}^{'} |c_j |.
\label{lab}
\end{equation}
The last term with coefficient $\lambda > 0$ is the newly-added regularization term ($\sum'$ indicates that the summation does not include the constant $E_c$), which penalizes the coupling constants $c_j$ to take a large value. In other words, minimizing the regularized cost function Eq.~\eqref{lab} is to decrease the error while keeping the coupling constants small. Then the gradient descent method leads to the solution where only a small number of parameters are essential. The $L_1$ norm regularization is particularly powerful and widely applied for sparse modeling problems where the number of parameters can be much larger than the number of data~\cite{Honma:2016aa,hukushima_2017,otsuki2017}. 

In Fig.~\ref{REGULARIZATION}, we show how the regularized cost function Eq.~\eqref{lab} works for the ansatz  Eq.~\eqref{REG_HAM}. Here we use the $L=10$ Hubbard chain with $U/t=8$ and take the $N=50$ low-energy eigenvalues in the $S^z = \sum_{i}^{L} S^z_i = 2$ sector for the optimization. Each parameter is normalized by its value without the regularization $\lambda = 0$. This plot tells us how sensitive these couplings are to the regularization. We see that $K_1$ and $M_i$ ($i = 1, ..., 6$) rapidly decrease as we increase the regularization strength $\lambda$ and for $\lambda > 10^{-6}$ the optimized Hamiltonian essentially contains only $J_1, J_2, J_3, K_2$, and $K_3$ (and the constant energy shift $E_c$) terms. The vanishing $K_1$ is consistent with the $O(t^5/U^5)$ perturbation calculations in Refs~\cite{PhysRevB.37.9753,Rej2006}. If we further increase the regularization strength the effective model changes from that at $O(t^5/U^5)$ to $O(t^3/U^3)$, and to $O(t/U)$. This demonstrates that the $L_1$ norm regularization is indeed useful for determining the ansatz. We can simply introduce various symmetry-allowed terms in the initial ansatz and use the gradient descent method for the regularized cost function. After obtaining the ansatz, we can perform the  optimization without the regularization to obtain the actual estimation of the effective model.

 \begin{figure}[htbp]
   \centering
\includegraphics[width = 120mm]{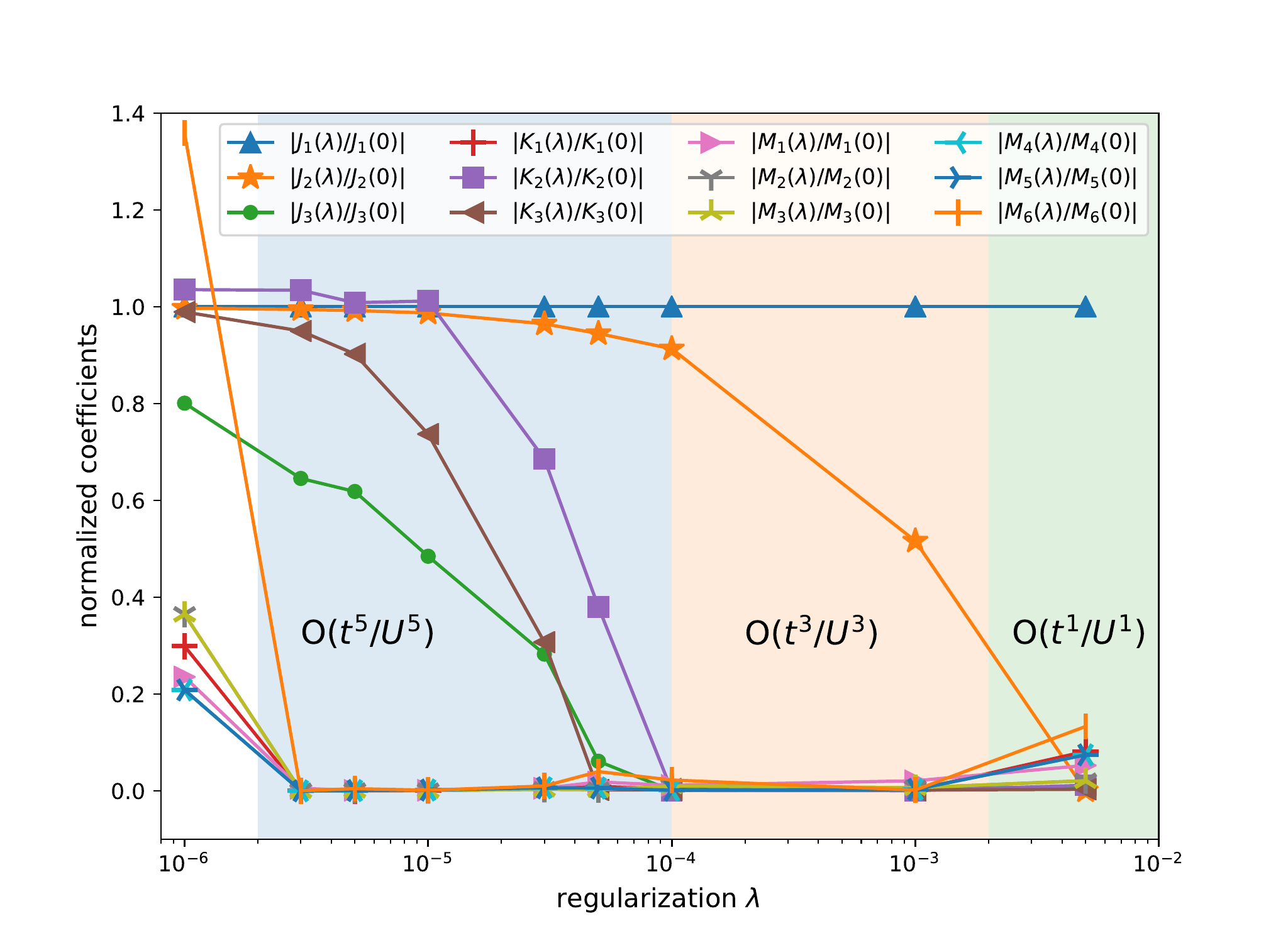}
       \caption{Optimized parameters of the model Eq.~\eqref{REG_HAM} with $L = 10$, $S^z =2$, $N=50$, $\alpha = 0.01$, and $U = 8$ for various regularization strength $\lambda$. Each coefficient is normalized by its estimated value for $\lambda = 0$. As we increase the regularization strength, coefficients corresponding to the high order terms in the perturbation theory drop systematically.}
       \label{REGULARIZATION}
  \end{figure} 

 \section{Stochastic optimization}
 If the cost function is a convex function globally, i.e. has no local minima, the simple optimization algorithm like the gradient descent method gives optimal parameters. However, as we have noted in the introductory part, energy spectrum of a Hamiltonian is a complicated, nonlinear function of coupling constants and the cost function is not a convex function generically. Nevertheless, as we demonstrate for the Hubbard model, if the ansatz is properly chosen the cost function can be made almost a convex one and we can achieve the global minimum (at least the one quite close to that). Therefore, one possible approach to the local minima problem is to apply the sparse modeling method discussed in appendix A and specify the proper form of the ansatz. Here, we introduce another approach, stochastic optimization for avoiding the deadlock to local minima. We review the stochastic gradient descent (SGD) method and its modern variant, Adam (Adaptive moment estimation), which is used in the last part of the main text.

 In the framework of the stochastic optimization, we first randomly reorder the dataset and pick up only one eigenvalue $E_j$. 
Then we update the parameters to decrease the value of $\frac{1}{2}(E_j^{\rm data} - E_j)^2$, or the cost defined from one spectrum, by the gradient descent method. After finishing the updates for all the data, we again randomly reorder the spectrum and repeat the same procedure. Because the parameters are optimized for a single datum at each step, not for the entire dataset, there is no guarantee that the cost defined for the full dataset decreases monotonically but on average it does. Due to the randomization, the SGD is more unlikely to be suffered from local minima of the cost function, i.e. has better performance for non-convex functions than the simple gradient descent method using all the data at once. Also we can improve the convergence of the SGD by using sophisticated optimizers such as Adam. 

Adam is an algorithm for the stochastic optimization problem designed to speed up the optimization process~\cite{Adam}. Adam requires four hyperparameters $\alpha, \beta_1, \beta_2, \epsilon$. The detail of the implementation can be found in the literature. In this paper, we use $\alpha = 0.001, \beta_1 = 0.99, \beta_2 = 0.999, \epsilon = 10^{-8}$, which are those recommended in the original paper of Adam~\cite{Adam}. We first obtain the $J_1$ using the conventional gradient descent method with the learning rate $0.001$ under the ansatz having only the nearest-neighbor Heisenberg interaction and use the obtained $J_1$ value as the initial value for the full ansatz Eq.~\eqref{HA} (initial values for other couplings $J_{2,...,5}, K_2,$ and $K_3$ are taken to be zero). 

 \section{Comment on the entanglement spectrum in the Haldane phase}
In the rung-singlet phases, the Heisenberg interaction in the rung direction is antiferromagnetic. Therefore, in the strong rung limit, the ground state is a product of spin singlets in the rung direction. Then the ES is flat and the EH is proportional to the identity matrix. On the contrary, when the rung interaction is attractive and dominant over the leg interaction, the ground state is a collection of spin triplets. As we have seen (and is observed in Ref.~\onlinecite{PhysRevB.83.245134}), though the ground state is a collection of independent spin triplets,  the ES is dispersive in this case because of the superposition of degenerate spin triplet states.
 
Let us consider a ladder with $L = 3$ (six sites in total). If we take the quantization axis of spins in the $z$ direction and represent the up and down spin states using arrows, spin triplet states for a pair of spins are represented as $\ket{\Psi_{-1}} = \ket{\downarrow_A \downarrow_B}$, $\ket{\Psi_{+1}} = \ket{\uparrow_A \uparrow_B}$, and $\ket{\Psi_{0}} = \frac{1}{\sqrt{2}}\left(\ket{\uparrow_A \downarrow_B}+\ket{\downarrow_A \uparrow_B}\right)$. In the strong rung limit, we have seven energetically degenerate eigenstates $\ket{\Psi}_{\pm} = \ket{\Psi_{0}} \ket{\Psi_{0}} \ket{\Psi_{0}} , \ket{\Psi_{0}} \ket{\Psi_{\pm1}} \ket{\Psi_{\mp1}}, \ket{\Psi_{\pm1}} \ket{\Psi_{\mp1}} \ket{\Psi_{0}}, \ket{\Psi_{\pm1}} \ket{\Psi_{0}} \ket{\Psi_{\mp1}} $ in the sector without total magnetization. If we take into account the translational invariance in the leg direction, only $\ket{\Psi} =  \ket{\Psi_{0}} \ket{\Psi_{0}} \ket{\Psi_{0}}$ is the momentum eigenstate as it is, and the others must form the superposition like $\ket{\Psi} = \frac{1}{\sqrt{3}}\left( \ket{\Psi_{0}} \ket{\Psi_{+1}} \ket{\Psi_{-1}}+ \ket{\Psi_{0}} \ket{\Psi_{+1}} \ket{\Psi_{-1}}+ \ket{\Psi_{-1}} \ket{\Psi_{0}} \ket{\Psi_{+1}}\right)$. 
 
If we trace out the spins in the $B$ leg, the state $\ket{\Psi}= \ket{\Psi_{0}} \ket{\Psi_{0}} \ket{\Psi_{0}}$, corresponding to the large-$D$ phase, yields the flat ES, while the others with the superposition of different spin triplets generate the dispersive ES.

}
\bibliography{References/ML_paper_references}
\end{document}